\documentclass[aps,prl,twocolumn,superscriptaddress]{revtex4}
\usepackage{graphicx}
\usepackage{amsmath,amssymb}

\begin{document}
\title{Controlling hysteresis in superconducting constrictions with a resistive shunt}
\author{Nikhil Kumar}
\affiliation{Department of Physics, Indian Institute of Technology Kanpur, Kanpur 208016, India}
\author{C. B. Winkelmann}
\affiliation{Universit\'e Grenoble Alpes, Institut N\'eel, F-38042 Grenoble, France}
\affiliation{CNRS, Institut N\'eel, F-38042 Grenoble, France}
\author{Sourav Biswas}
\affiliation{Department of Physics, Indian Institute of Technology Kanpur, Kanpur 208016, India}
\author{H. Courtois}
\affiliation{Universit\'e Grenoble Alpes, Institut N\'eel, F-38042 Grenoble, France}
\affiliation{CNRS, Institut N\'eel, F-38042 Grenoble, France}
\author{Anjan K. Gupta}
\affiliation{Department of Physics, Indian Institute of Technology Kanpur, Kanpur 208016, India}
\date{\today}

\begin{abstract}
We demonstrate control of the thermal hysteresis in superconducting constrictions by adding a resistive shunt. In order to prevent thermal relaxation oscillations, the shunt resistor is placed in close vicinity of the constriction, making the inductive current-switching time smaller than the thermal equilibration time. We investigate the current-voltage characteristics of the same constriction with and without the shunt-resistor. The widening of the hysteresis-free temperature range is explained on the basis of a simple model.
\end{abstract}
\maketitle

\section{Introduction}
A superconducting weak-link (WL), such as a constriction, between two bulk superconductors is of interest for its Josephson junction-like properties and subsequent application to micron size superconducting quantum interference devices ($\mu$-SQUIDs) \cite{tinkham-book,likharev-rmp}. The latter can be used in probing magnetism at small scales \cite{mic-squid-appl,klaus-veauvy-physica,vasyukov-nature,hao-apl}. Hysteresis present in current-voltage characteristics (IVCs) is a limiting factor in WL-based SQUIDs. In a hysteretic IVC when the current is ramped up from zero, the device typically switches to a non-zero voltage state at the critical current $I_c$. The subsequent current ramp-down gives a switching to zero-voltage state at a smaller current, called re-trapping current $I_r$. Hysteresis in IVCs is seen at low temperatures and disappears above a crossover temperature $T_h$ as $I_c$ and $I_r$ meet \cite{hazra-prb,nikhil-akg-proximity,JAP-cooling-fin}. In a conventional tunnel-barrier type Josephson-junction, hysteresis arises from large junction capacitance  and can be eliminated by adding a shunt-resistor in parallel to the junction \cite{tinkham-book,squid-sensitivity}. The effect of the shunt resistor on nano-wire based WL devices was modeled recently using resistively and capacitively shunted junction (RCSJ) model with an effective capacitive time \cite{bezryadin-prb}. The hysteresis in similar devices is well understood using the thermal model \cite{tinkham-prb}. The hysteresis in WLs is due to local Joule-heating \cite{skocpol-jap,herve-prl}, which gives rise to a self-sustained resistive hot-spot in the WL region, even below $I_c$.

Eliminating thermal hysteresis in WLs has been the subject of intense research in the past years. A normal metal shunt directly on top of the constriction \cite{lam-apl, dibyendu-apl1,dibyendu-apl2} has been tried, but it affects both the superconductivity and thermal properties in a way that depends on the interface transparency. Using a bilayer with a superconductor (that can locally etched with a Focussed Ion Beam) covering a normal metal film allows one to obtain a WL that is also a good thermal bath \cite{moseley-apl}. A parallel shunt resistor far away from the WL \cite{muck-apa} is more flexible approach, but it gives rise to relaxation oscillations due to large inductive time for switching of the current between the WL and the shunt. The performance of such SQUIDs with a distant shunt-resistor is eventually similar to that of the hysteretic ones \cite{mic-squid-appl,muck-apa}. A systematic study of the ability of a parallel shunt in preventing both the thermal runaway and hysteresis is thus highly desirable.

\begin{figure}
\begin {center}
\includegraphics[width=8cm]{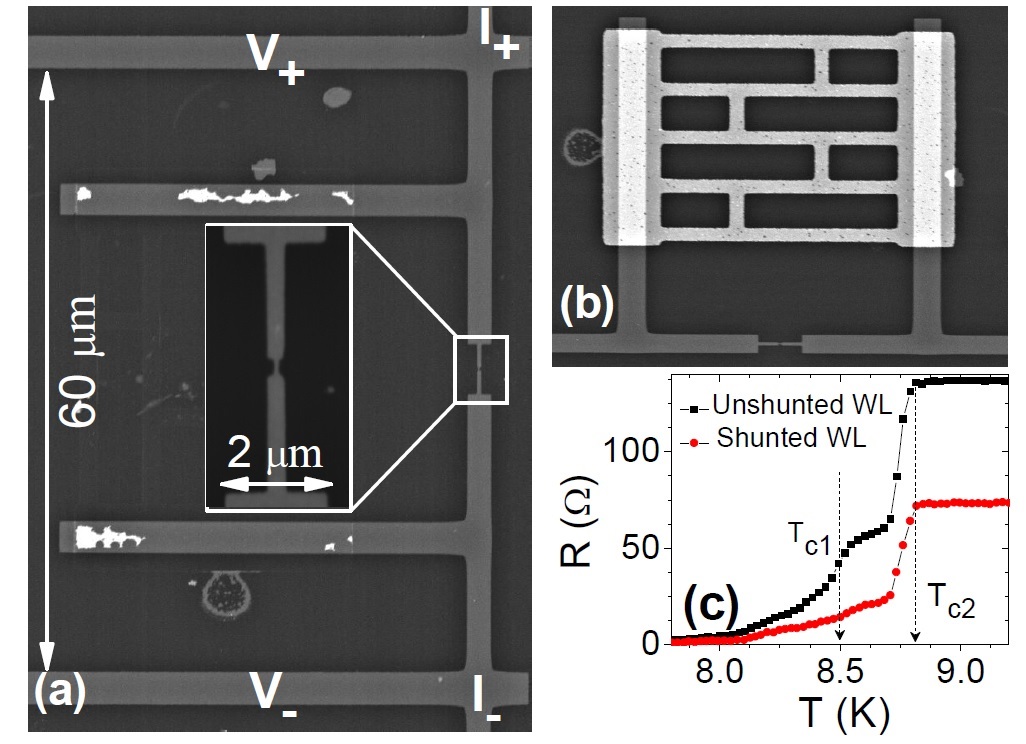}
\caption{(a) SEM image of the WL device after etching the gold shunt. The inset in (a) shows the zoomed-in SEM image of the WL with patterned width 70 nm and length 150 nm. (b) SEM image of the same device rotated by $90^\circ$, before etching the hatch pattern shunt resistance made of gold microwires. (c) Resistance vs temperature (at 0.01 mA bias current) for the shunted and unshunted WLs, showing different transitions at $T_{c1}$ and $T_{c2}$ respectively corresponding to the narrow and the wide leads.}
\label{schema}
\end{center}
\end{figure}

The role of a shunt-resistor can be understood using a simple quasi-static thermal model discussed by Tinkham et al. \cite{tinkham-prb}. In this model, the heat generated in the resistive hot-spot in a long constriction is conducted (only) through electronic conduction to the large electrodes at the end. The thermal conductivity $K$ of the normal metal and superconductor are assumed to be identical and temperature-independent. The re-trapping current is then found to be $I_r(T_b)=I_r(0)\sqrt{1-T_b/T_{c}}$ with $I_r(0)=4\sqrt{KAT_{c}/LR_n}$. Here $R_n$ is the normal resistance of the constriction of length $L$ and cross-sectional area $A$, $T_b$ is the bath temperature and $T_{c}$ is the superconductor critical temperature. From Ginzburg-Landau theory \cite{likharev-rmp} the critical current follows $I_c(T_b)=I_{c}(0)(1-T_b/T_c)$ in the regime $T_b>T_c/2$. Thus $I_c$ and $I_{r}$ cross at a crossover temperature $T_{h}=T_c[1-[I_{r}(0)/I_c(0)]^2]$ \cite{Th-reduction}. In the presence of a shunt resistor $R_s$, the bias current is shared between the shunt and the WL, when the latter is resistive. Thus, in $I_r(0)$ expression, $1/R_n$ is replaced by $(1/R_n)+(1/R_s)$. As a result $I_r(0)$ changes to a higher value given by \begin{eqnarray}I_{rs}(0)=I_{r}(0)\sqrt{1+R_n/R_s}.\label{eq-1}\end{eqnarray} In contrast, $I_{c}$ remains unaffected. Hence the crossover temperature decreases and the hysteresis-free temperature range $[T_{h},T_c]$ widens thanks to the shunt. For eliminating hysteresis above temperature T, $R_s$ with value less than $R_{sc}=R_n/[\{I_{c}(T)/I_{r}(T)\}^2-1]$ will be required. The assumed immediate sharing of the bias current between WL and $R_s$ implies a small inductive current-switching time as compared to the thermal equilibration time. The minimum shunt resistor value, $R_{sc}$, from our simple model can also describe the behavior of shunted nano-wire devices studied by Brenner et al. \cite{bezryadin-prb}.

In this Letter, we compare the current-voltage characteristics of carefully designed WL devices with (and without) a shunt resistor kept in close vicinity of the WL, thus making the inductive current-switching time smaller than the thermal equilibration time. We observe an increase in the re-trapping current and a widening of the hysteresis-free temperature range thanks to the shunt, which we discuss using the above model.

\section{Experimental details}
Devices were fabricated on Si substrates in two subsequent lithography and e-beam deposition steps as follows: 1) laser lithography of hatch patterned shunt resistor \cite{shunt-design} and alignment marks on a photo-resist, 2) deposition and lift-off of a Ti (3 nm)/Au (20 nm) layer, 3) oxygen reactive-ion-etching (RIE) to remove residual resist, 4) deposition of a 31 nm thick Nb-film, 5) aligned electron beam lithography of a PMMA resist of the WL pattern, 6) deposition and lift-off of a 20 nm thick Al-film, 7) etch of Nb with SF$_6$-RIE, 8) chemical removal of Al. After fully characterizing the shunted device, the Au shunt was etched using a KI-I$_2$ solution which does not attack Niobium. Electrical transport studies down to 1.3 K were pursued using a closed cycle He-refrigerator \cite{ice-CCR} with a homemade sample holder that incorporates copper powder filters. The data were recorded using data acquisition cards and homemade analog electronics. Two nominally identical devices demonstrated similar results.
\begin{figure}
\begin {center}
\includegraphics[width=8cm]{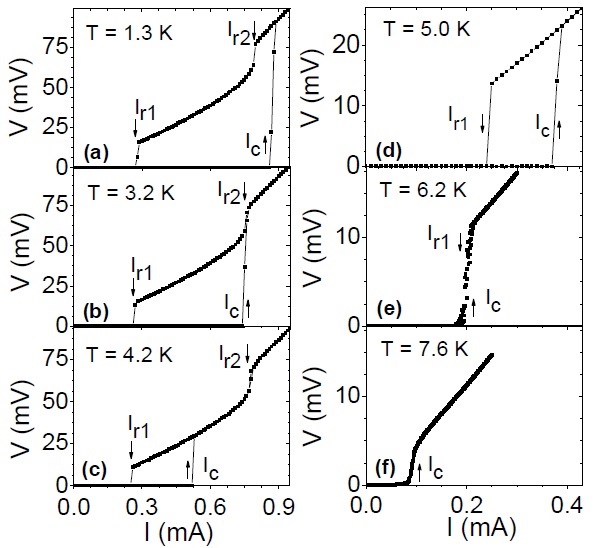}
\caption{(a) I-V characteristics for the unshunted weak link at a set of different temperatures.}
\label{IVCUnshunted}
\end{center}
\end{figure}

\section{Results and Discussions}
Fig. \ref{schema} shows the electron micrograph and resistance vs temperature of the reported device. The WL [see inset of Fig. \ref{schema}(a)] as designed has a 150 nm length and a 70 nm width. Narrow leads (width 0.3 $\mu$m, length 2.4 $\mu$m) with normal resistance $2R_1$, connect symmetrically to the two ends of the WL of normal resistance $R_{WL}$. Wide leads of width 2 $\mu$m connect the narrow leads to the shunt and the voltage probes. Fig. \ref{schema}(c) shows the measured resistance $R$ versus temperature $T$ before [see Fig. \ref{schema}(b)] and after [see Fig. \ref{schema}(a)] etching the Au shunt. For the two cases, $R$ drops from a saturation value of 135 or 72 $\Omega$ (at 10 K) with two transitions at $T_{c1}$ = 8.5 K and $T_{c2}$ = 8.8 K. These two critical temperatures are attributed respectively to the narrow-leads and the wide leads. Let us stress that the critical temperatures $T_{c1,2}$ are not affected by the removal of the Au shunt, which confirms that the shunt etching did not damage the niobium pattern. From the resistance drop at $T_{c2}$ for the unshunted device, we find a square resistance $R_{\Box}$ = 3.5 $\Omega$ giving a resistivity value of 10.8 $\mu\Omega$.cm for the Nb film. From the narrow lead resistance and dimensions, we estimate $R_{WL}$ = 9 $\Omega$ and $2R_1$ = 56 $\Omega$. By comparing the resistances of the two devices just below $T_{c2}$, we find $R_s$ = 35.6 $\Omega$, which is consistent with separate measurements of Au wires' resistances.
\begin{figure}
\begin {center}
\includegraphics[width=8cm]{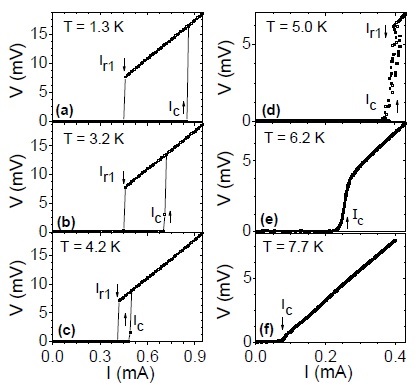}
\caption{(a) I-V characteristics for the same weak link device with a parallel shunt, at the same set of different temperatures as in Fig. \ref{IVCUnshunted}.}
\label{IVCShunted}
\end{center}
\end{figure}

While designing the device, we have kept the shunt-resistor close to the WL to minimize the associated loop-inductance $L_{sh}$. The inductance of a square loop with sides $a$ and width $b \ll a$, is given by $\frac{2}{\pi}\mu_0a \ln[\frac{2a}{b}]$ \cite{grower-inductance-calculations}. Using this relation with $a$ = 20 $\mu$m and b = 3 $\mu$m, we estimate the inductance of the loop containing gold shunt hatch pattern and the Nb leads as $L_{sh}$ = 40 pH. This gives a inductive-current-switching-time $\tau_L = L_{sh}/(R_s+R_{WL}) \simeq$ 1 ps. The heat is transferred to the substrate over a length-scale given by thermal healing length $l_{th} (= \sqrt{Kt/\alpha}=1.6 \mu$m) \cite{nikhil-akg-proximity,supp-mat}. Here $t$ is the film thickness and $\alpha$ is the interface heat loss coefficient. Thus the thermal cooling time is $\tau_T={l_{th}}^2/{{\pi}^2D}=ct/\pi^2\alpha$ with D (=$K/c=1$ cm$^2$/s) as the diffusion constant and $c$ as the volumetric heat capacity. We thus estimate the thermal time $\tau_T$ = 2.5 ns, which is much larger than $\tau_L$. When the WL switches from the superconducting to the resistive state, the current redistribution between the shunt and the WL is thus much faster than the thermal runaway in the device.

Figure \ref{IVCUnshunted} shows IVCs of the unshunted device at various temperatures. At low temperatures, sharp voltage jumps and drops are observed at the critical current $I_c$ and at two re-trapping currents $I_{r1}$ and $I_{r2}$. The latter two arise from thermal instabilities respectively in the WL plus the narrow leads ($I_{r1}$) and in the wide leads ($I_{r2}$) \cite{nikhil-akg-proximity}. At 1.3 K, $I_{c}$ is higher than both $I_{r1}$ and $I_{r2}$ [see Fig. \ref{IVCUnshunted}(a)]. The IVC slope above $I_{r2}$ is 142 $\Omega$, slightly larger than the normal state value of 135 $\Omega$ because of over-heating. At higher temperatures when $I_{r2}$ is smaller and heating less, the slope is 135 $\Omega$. The IVC slope above $I_{r1}$ is 73 $\Omega$, which is close to the combined resistance $R_{WL} + 2R_1$, i.e. 65 $\Omega$. The slightly larger value is due to the spread of the hot-spot into the wide leads. With increasing temperature, $I_c$ crosses $I_{r2}$ near 3.2 K [see Fig. \ref{IVCUnshunted}(b)] and it merges with $I_{r1}$ near $T_h$ = 6.25 K [see Fig. \ref{IVCUnshunted}(e)]. At higher temperature, the IVC is non-hysteretic and the resistance for $I>I_c$ is 65 $\Omega$, indicating that the WL as well as the narrow leads are resistive.

Figure \ref{IVCShunted} shows IVCs of the same device but prior to the shunt removal. We observe voltage jumps and drops at $I_c$ and $I_{r1}$ while the second retrapping current $I_{r2}$ is visible only in IVCs with a larger bias current excursion \cite{supp-mat}. In the resistive region, the slope is always 22 $\Omega$ which corresponds to the parallel combination of the normal resistance of the WL plus the narrow-leads with the shunt, \mbox{i.e.} $(R_s^{-1}+(2R_1+R_{WL})^{-1})^{-1}$. The critical current $I_c$ magnitude at low temperatures is the same, within the error bars, as that of the unshunted device, confirming that the shunt removal did not damage the WL. Remarkably, the re-trapping current $I_{r1}$ has a higher value as compared to that of the unshunted device. As a result of $I_{r1}$ enhancement, $I_c$ and $I_{r1}$ meet at a lower crossover temperature $T_{hs}$ = 5 K in the shunted device, see Fig. \ref{IVCShunted}(d).

\begin{figure}
\begin {center}
\includegraphics[width=8cm]{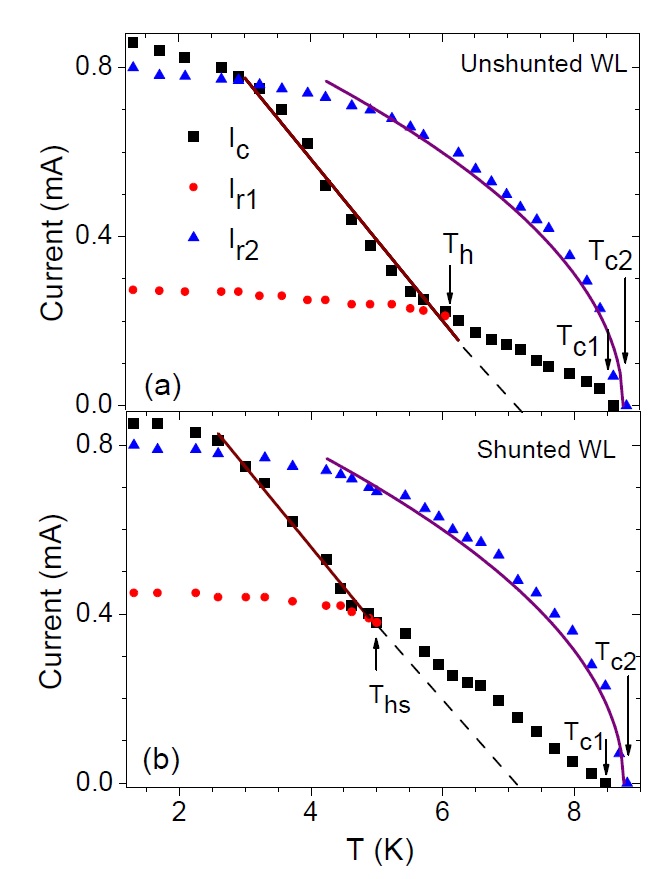}
\caption{Temperature evolution of $I_c$, $I_{r1}$ and $I_{r2}$ for (a) an unshunted weak link (WL) and (b) the same WL with a parallel shunt resistor, showing the reduction of crossover temperature in shunted case. The symbols are the data points. The continuous lines are the fits given by (in mA and K) $I_c(T_b) = 0.18 (7.2-T_b)$ and $I_{r2}(T_b)= 0.32(T_{c2}-T_b)^{1/2}$ for both the devices.}
\label{IcTWL}
\end{center}
\end{figure}

We summarize the temperature dependence of $I_c$, $I_{r1}$ and $I_{r2}$ for both devices in Fig. \ref{IcTWL}. In every case, the retrapping current $I_{r2}$ nearly follows a square root dependence with the bath temperature \cite{nikhil-akg-proximity} extrapolating to zero at $T_{c2}$. This is consistent with $I_{r2}$ being related to the thermal instability of the wide leads. Between about 3 K and the crossover temperature $T_h$ or $T_{hs}$, the critical current $I_c$ of both devices decreases linearly with temperature. The extrapolated critical temperature $T_c$ close to 7.2 K is that of the WL itself. In both the devices, the critical current $I_c$ decays markedly slower above $T_h$ or $T_{hs}$, owing to the proximity effect \cite{nikhil-akg-proximity}.

The retrapping current $I_{r1}$ follows a similar temperature dependence in the two devices but with a higher magnitude, by a factor of about 1.64, in the shunted device. This factor is similar to that found from Eq. \ref{eq-1}, i.e. 1.67, by using $R_s=$ 35.6 $\Omega$ and $R_n=2R_1+ R_{WL}=$ 65 $\Omega$. A more appropriate model, incorporating the interface heat loss, for our device configuration provides a similar agreement, where the ratio is found to be close to 1.60 \cite{supp-mat}. For our devices, both the models give similar agreement \cite{supp-mat}, as the conduction dominates over interface heat loss because the narrow leads' length is comparable to $l_{th}$ \cite{nikhil-akg-proximity}. The two models have significant disagreement when the constriction is much longer than $l_{th}$ \cite{supp-mat}. Thanks to the shunt and the related $I_{r1}$ enhancement, the hysteresis-free temperature range has increased from [6.25 K, 8.6 K] to [5 K, 8.6 K], see Fig. \ref{IcTWL}. For instance, the WL without shunt is hysteretic at 5 K, see Fig. \ref{IVCUnshunted}(d), while the one with shunt is non-hysteretic, see Fig. \ref{IVCShunted}(d).

The merging of $I_c$ and $I_{r1}$ above $T_h$ is different from our earlier results on unshunted $\mu$-SQUIDs \cite{nikhil-akg-proximity} where a crossing of the two was seen at $T_h$. Due to the presence of the SQUID loop, the heat evacuation in the $\mu$-SQUIDs is more efficient. For single WL devices with similar $I_c$ values, the less efficient heat evacuation favors merging over crossing. In fact, just below $T_c$, where $I_c$ is small, heat evacuation eventually dominates, and we do see distinct signatures of both $I_c$ and $I_{r1}$ in IVCs. We also see large fluctuations in voltage close to $T_h$ and for currents near $I_c$ in both the devices, see Fig. \ref{IVCUnshunted}(e) and \ref{IVCShunted}(d). From the time-series data we find a bistable telegraphic-like voltage signal in this regime. Thus time averaged voltages in the IVCs show significant fluctuations. Close to the boundary of the bistable regime, more sensitivity to noise is indeed expected.

\begin{figure}
\begin {center}
\includegraphics[width=8cm]{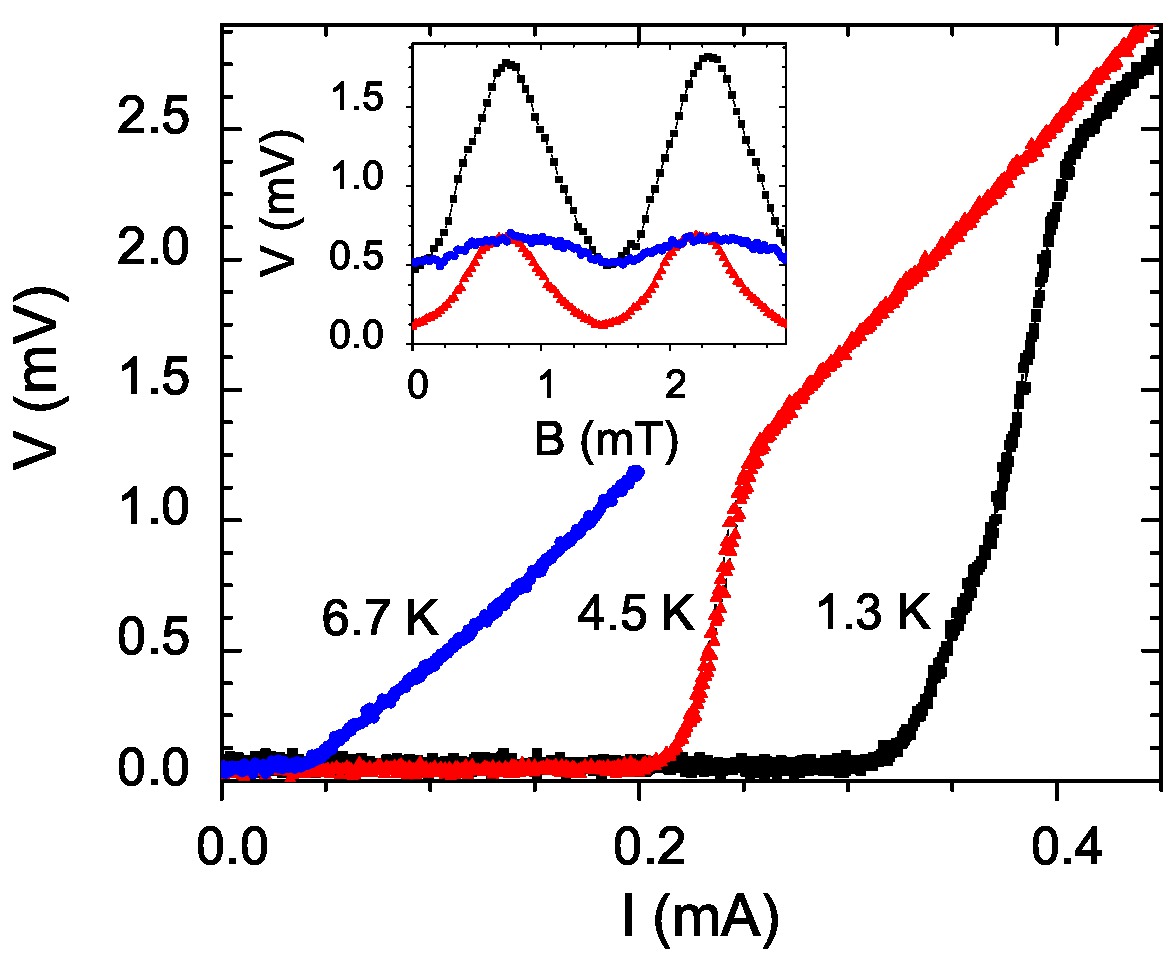}
\caption{IVCs of a shunted $\mu$-SQUID at 1.3, 4.5, and 6.7 K showing non-hysteretic characteristics. The inset shows the SQUID oscillations at similar temperatures biased close to the critical current namely 0.37 mA for 1.3 K, 0.22 mA for 4.5K and 0.05 mA for 6.7 K. The vertical scale at 6.7 K has been enhanced by a factor of 5 for better clarity in large scales.}
\label{IVCSQ}
\end{center}
\end{figure}

We have also studied a shunted $\mu$-SQUID device, with the same shunt geometry and resistance $R_s$ value as in the single WLs discussed above. Although the critical current $I_c$ is smaller in this SQUID  due to the reduced width ($< 50 nm$) of the weak links. The SQUID loop pattern is otherwise identical to our earlier work \cite{nikhil-akg-proximity}. Figure \ref{IVCSQ} shows the IVCs of the shunted $\mu$-SQUID at different temperatures, which is found to be non-hysteretic down to 1.3 K temperature, establishing the role of the shunt in widening the non-hysteretic temperature range for both WLs and $\mu$-SQUIDs. SQUID oscillations are clearly observed, see Fig. \ref{IVCSQ} inset.

Finally, let us discuss how we could further increase the re-trapping current and hence expand the hysteresis-free temperature range. Using a lower $R_s$ value will increase $I_{r1}$ further and widen the hysteresis-free temperature range for a given WL device. Nevertheless, this will also reduce the overall normal resistance and result in a lower voltage signal to be measured. The same can also be achieved by using a smaller $I_c$ WL with same $R_s$ value. We have verified this claim in another shunted WL device, with same $R_s$ value and smaller $I_c$, showing a $T_{hs}$ below 4.2 K. In any case, the shunt-resistor has to be kept close enough to the WL, so as to avoid relaxation oscillations, but not too close to cause heat or electron sharing between WL and the shunt, which can affect the WL superconductivity.

\section{Conclusions}
In summary, we have demonstrated a significant improvement of the hysteretic behavior of superconducting WLs and $\mu$-SQUIDs, using a parallel resistive shunt in close vicinity to the WL. As a result of the shunt, the hysteresis-free temperature range is wider. Our results can help to further develop WL-based non-hysteretic devices such as SQUIDs.

\section*{Acknowledgements}
Samples were fabricated at the platform Nanofab, CNRS Grenoble and measurements were carried out in IIT Kanpur. AKG thanks University Joseph Fourier for a visiting fellowship. NK acknowledges the financial support from CSIR, India. This work has been financed by the French Research National Agency, ANR-NanoQuartet (ANR12BS1000701) and the CSIR of the govt. of India.

\section{Supplementary Material}

\section{Hot Spot Model for the shunted Weak Links}

We discuss here a more elaborate quasi-static thermal model for our devices. This is a special case of the model discussed by Skocpol et. al. \cite{skocpol-jap}. In our devices, the WL is quite short, as compared to thermal length, and the normal metal-superconductor (NS) interface occurs in the adjacent narrow leads which loose heat to the bath via the interface with the substrate. \begin{figure}
\includegraphics[width=8cm]{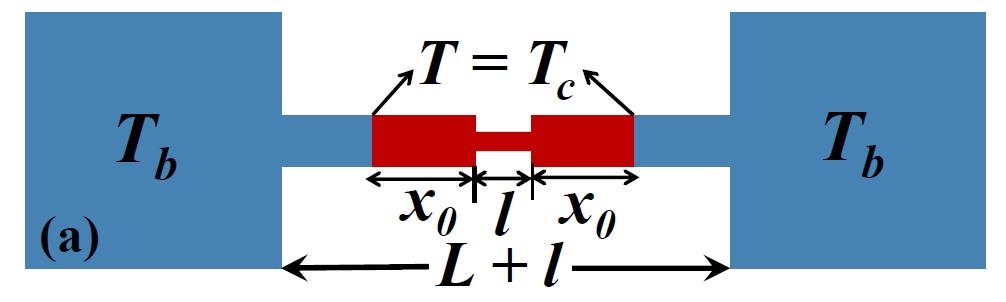}
\includegraphics[width=8cm]{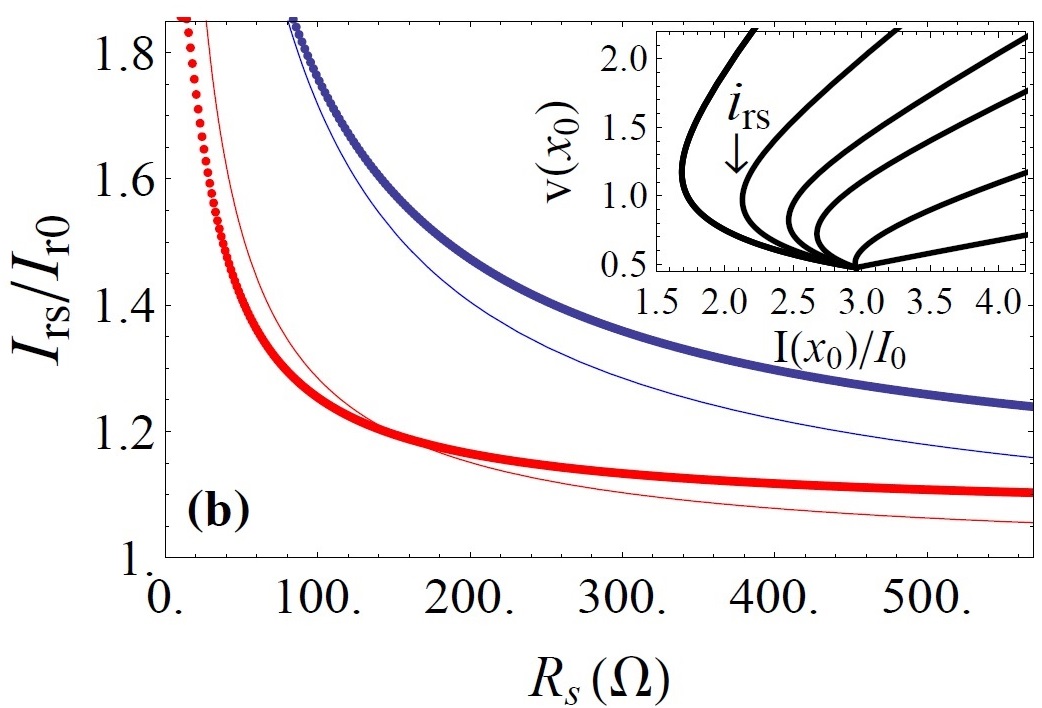}
\caption{(a) Schematic of the device showing the hot-spot in narrow leads, while the short weak link of length $l$ act as a power source. (b) $I_{rs}/I_{r0}$ variation with $R_s$ as found from the generalized thermal model (symbols) and simple thermal model (thin continuous lines). These plots are shown for two different lengths, $L=3 l_t$ (red curve) and $L=9 l_t$ (blue curve), of the narrow leads. The inset in (b) shows the normalized IVCs for $L=3 l_t$ with $R_s$ values of $\infty$, 100 (marked with the arrow), 50, 35.6, 20 and 10 $\Omega$. Here, $v(x_0)=V(x_0)/2I_0R_1$.}
\label{thermalmodel}
\end{figure}The heat conduction, together with the interface heat loss, occurs in the narrow leads (of length $L$, width $w$ and thickness $t$) that join the wide leads as shown in Fig. \ref{thermalmodel}(a). We assume that the temperature where the narrow lead meets the wide one is $T_b$ since the wide-lead is much wider than the narrow one. We treat the WL as a simple source of $I^2R_{WL}$ heat when it is resistive. The unsustainability of NS interface in the narrow lead determines the magnitude of the re-trapping current. Near the NS interface, the temperature is close to $T_c$ and thus the thermal conductivity ($K$) does not have much variation near this interface. $K$ is then considered as a constant. Solving the heat flow equations for the narrow leads, we get the relation between bias current and the NS interface location, $x_0$, as $I=I_0 i(x_0)$ with
\begin{equation}
I_0=\sqrt{\frac{\alpha w^2t(T_c-T_b)}{\rho_n}}.
\label{eq:i-x0}
\end{equation}
and
\begin{equation}
i(x)=\sqrt{\frac{\sinh\left(\frac{x}{l_{t}}\right)[1 + \coth\left(\frac{x}{l_{t}}\right) \coth\left(\frac{L}{2l_{t}}-\frac{x}{l_{t}}\right)]}{q + \sinh\left(\frac{x}{l_{t}}\right)}}.
\label{eq:i-x1}
\end{equation}

Here, $\rho_n$ is the the normal state resistivity of the narrow lead. $l_{t}=\sqrt{Kt/\alpha}$ is the thermal healing length, which is defined as the length scale over which the heat ($\dot{Q_0}=w\alpha l_{t}(T_c-T_b)$) is transferred to the substrate. Here the coefficient $\alpha$ is characteristic of the interface. $q$ represents the ratio of $R_{WL}$ to the resistance of $2l_{t}$ length of the narrow lead. The voltage across the WL for a given $x_0$ is: $V(x_0) = I_0 i(x_0)[4R_1(x_0/L)+R_{WL}]$. In the presence of $R_s$ the bias current is given by $I=I_0i(x_0)+[V(x_0)/R_s]$, which gives a different $x_0$ value, while the voltage remains the same. The inset of Fig. \ref{thermalmodel}(b) shows the calculated IVCs for different $R_s$ values with a current minimum at $i_{rs}=I_{rs}/I_0$, which defines the normalized re-trapping current. Further, $i_{rs}$ increases with reducing $R_s$ value, see Fig. \ref{thermalmodel}(b). We find that $i_{rs}$ value mainly depends on the narrow-lead and shunt resistance values but it does not depend much on $R_{WL}$. $I_0$ follows a square root dependence on $(T_c-T_b)$ implying the same for $I_{rs}$. The increment in re-trapping current by the parallel shunt resistance is in agreement with our model with $R_s=$ 35.6 $\Omega$, $R_n = 2R_1 =$ 56 $\Omega$ and $R_{WL}=$ 9 $\Omega$.

Fig. \ref{thermalmodel}(b) shows the enhancement of retrapping current for two different lengths of the narrow lead. For comparison, we have also plotted results from the simple model (Eq. 1 of the main manuscript). Since the length of narrow leads (L=3 $l_{t}$) in our devices is comparable to $l_{t}$, we are still in a regime, where conduction dominates over surface heat loss. Thus both the models give similar enhancement in retrapping current. With longer length of the narrow leads, the simple model deviates significantly from the detailed model as seen in Fig. \ref{thermalmodel}(b) for $L=9 l_t$.

\begin{figure}
\includegraphics[width=7cm]{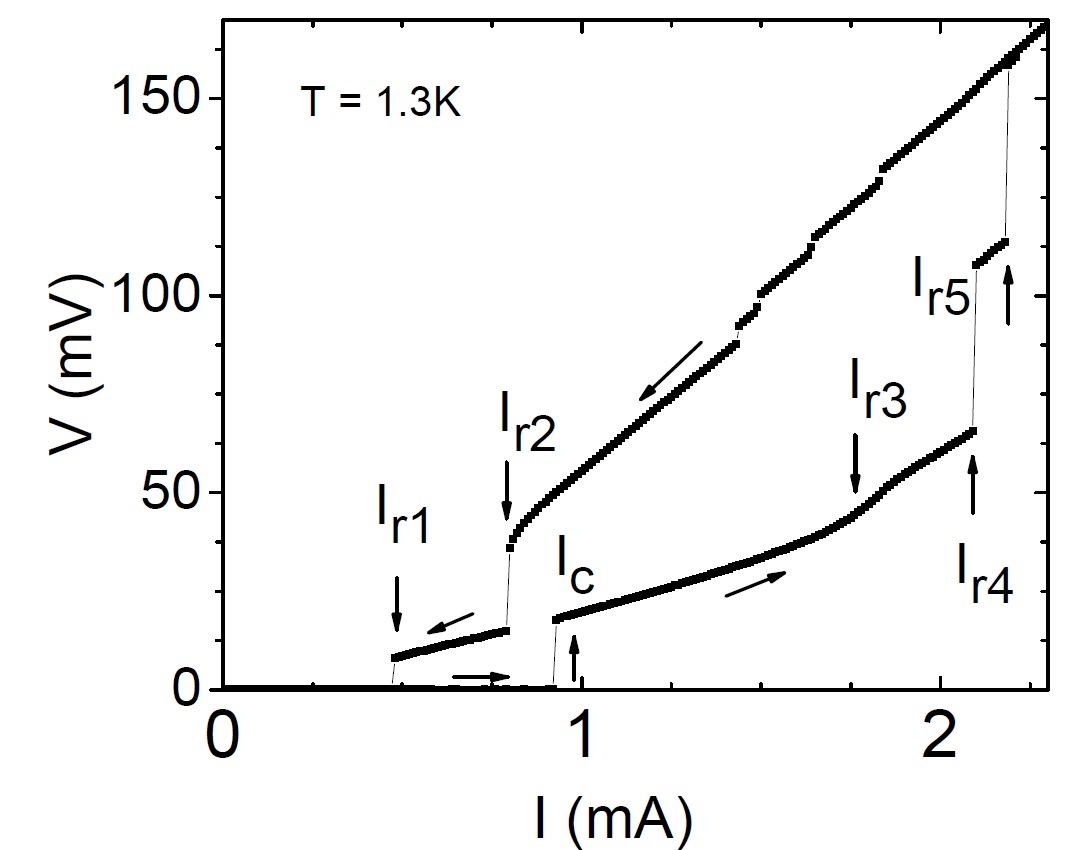}
 \caption{IVC at higher bias currents (as an extension to Fig. 3 of the main text) at 1.3 K temperature, for shunted WL. This shows the different transitions corresponding to the thermal runaway in various portions of the device.}
 \label{IVCrunaway}
\end{figure}

\section{IVC of the shunted device at high bias currents}

Fig. \ref{IVCrunaway} shows IVCs at higher bias currents, where we see several jumps in voltage occurring at different currents. At 1.3 K these currents are: $I_{r1}$ = 0.45 mA, $I_{r2}$ = 0.78 mA, $I_c$ = 0.85 mA, $I_{r3}$ = 1.80 mA, $I_{r4}$ = 2.08 mA, and $I_{r5}$ = 2.18 mA. We see several kinks between 1.82 mA and 1.43 mA while ramping down the current. The jumps at $I_c$, $I_{r3}$, $I_{r4}$ and $I_{r5}$ are seen, when we ramp the current up, while the others occur during current ramp down. These transitions have been seen reproducibly in three different devices and are attributed to the thermal instability of N-S interface in different superconducting portions. The magnitude of $I_{r2}$ agrees with the instability current in 2 $\mu$m wide leads \cite{nikhil-akg-proximity}. All the five re-trapping currents follow similar square-root dependence on $(T_c-T_b)$ as expected for the instability currents \cite{nikhil-akg-proximity}.

\section*{References}
\begin{thebibliography}{10}
\bibitem{tinkham-book}  M. Tinkham, Introduction to Superconductivity 2nd ed. (Mc Graw-Hill, New York, 1996).
\bibitem{likharev-rmp}  K. K. Likharev, Rev. Mod. Phys. {\bf 51}, 101 (1979).
\bibitem{mic-squid-appl} W. Wernsdorfer, Adv. Chem. Phys. {\bf 118}, 99 (2001).
\bibitem{klaus-veauvy-physica}  K. Hasselbach, C. Veauvy, D. Mailly, Physica C {\bf 332}, 140 (2000).
\bibitem{vasyukov-nature} D. Vasyukov, Y. Anahory, L. Embon, D. Halbertal, J. Cuppens, L. Neeman, A. Finkler, Y. Segev, Y. Myasoedov, M. L. Rappaport, M. E. Huber, and E. Zeldov, Nature Nanotech. {\bf 8}, 639 (2013).
\bibitem{hao-apl} L. Hao, J. C. Macfarlane, J. C. Gallop, D. Cox, J.Beyer, D. Drung, and T. Schurig, Appl. Phys. Lett. {\bf 92}, 192507 (2008).
\bibitem{hazra-prb} D. Hazra, L. M. A. Pascal, H. Courtois, and A. K. Gupta, Phys. Rev. B {\bf 82}, 184530 (2010).
\bibitem{nikhil-akg-proximity} Nikhil Kumar, T. Fournier, H. Courtois, C. B. Winkelmann, and A. K. Gupta, Phys. Rev. Lett. {\bf 114}, 157003 (2015).
\bibitem{JAP-cooling-fin} A. Blois, S. Rozhko, L. Hao, J. C. Gallop, and E. J. Romans, J. Appl. Phys. {\bf 114}, 233907 (2013).
\bibitem{squid-sensitivity} J. Clarke and A. I. Braginski, The SQUID Handbook, Vol. 1, WILEY-VCH Verlag GmbH and Co. KGaA, 2004.
\bibitem{bezryadin-prb} Matthew W. Brenner, Dibyendu Roy, Nayana Shah, and Alexey Bezryadin, Phys. Rev. B {\bf 85}, 224507 (2012).
\bibitem{tinkham-prb} M. Tinkham, J. U. Free, C. N. Lau, and N. Markovic, Phys. Rev. B {\bf 68}, 134515 (2003).
\bibitem{skocpol-jap} W. J. Skocpol, M. R. Beasley, and M. Tinkham, J. Appl. Phys. {\bf 45}, 4054 (1974).
\bibitem{herve-prl} H. Courtois, M. Meschke, J. T. Peltonen, and J. P. Pekola, Phys. Rev. Lett. {\bf 101}, 067002 (2008).
\bibitem{lam-apl} S. K. H. Lam and D. L. Tilbrook, Appl. Phys. Lett. {\bf 82}, 1078 (2003).
\bibitem{dibyendu-apl1} D. Hazra, J. R. Kirtley, and K. Hasselbach, Appl. Phys. Lett. {\bf 103}, 093109 (2013).
\bibitem{dibyendu-apl2} D. Hazra, J. R. Kirtley, and K. Hasselbach, Appl. Phys. Lett. {\bf 104}, 152603 (2014).
\bibitem{moseley-apl} R. W. Moseley, W. E. Booij, E. J. Tarte, and M. G. Blamire, Appl. Phys. Lett. {\bf 75}, 262 (1999).
\bibitem{muck-apa} M. M\"{u}ck, H. Rogalla, and C. Heiden, Appl. Phys. A {\bf 47}, 285-289 (1988).
\bibitem{Th-reduction} This assumes the $T_c$ of the WL to be the same as the adjecent superconducting structure, which is actually not true for our devices. $T_h$ will get further reduced, widening the hysteresis free temperature range due to different $T_c$'s corresponding to $I_c$ and $I_{r1}$, owing to proximity effect \cite{nikhil-akg-proximity}.
\bibitem{shunt-design} This topology of the shunt was used with a motive to study how the shunt resistance value changes the IVCs of the WL. Thus we had planned to cut various struts of the shunt using focussed ion beam (FIB) to modify the resistance value. However, the yield in this process was extremely poor and thus we decided to remove the shunt altogether.
\bibitem{ice-CCR} ICE OXFORD made Closed Cycle Refrigerator.
\bibitem{grower-inductance-calculations} F. W. Grover, Inductance Calculations: Working Formulas and Tables, Dover Publications Inc., New YORK (1962).
\bibitem{supp-mat} See the supplementary material for further details.
\end{thebibliography}
\end{document}